\documentclass[a4paper,english,american]{article}
\usepackage[T1]{fontenc}
\usepackage[latin9]{inputenc}
\usepackage{amsmath}
\usepackage{graphicx}


\usepackage{babel}

\begin{document}

\title{CaSPA - an Algorithm for Calculation of the Size of Percolating Aggregates}

\author{James E. Magee%
\thanks{\selectlanguage{english}%
\noindent Corresponding author.\protect \\
E-Mail: j.magee@manchester.ac.uk\selectlanguage{american}
}, Helen Dutton, Flor R. Siperstein\\
 School of Chemical Engineering and Analytical Science,\\
The University of Manchester, PO Box 88,\\
Sackville Street, Manchester, M60 1QD, United Kingdom}

\maketitle
\begin{abstract}
We present an algorithm (CaSPA) which accounts for the effects of
periodic boundary conditions in the calculation of size of percolating
aggregated clusters. The algorithm calculates the gyration tensor,
allowing for a mixture of infinite (macroscale) and finite (microscale)
principle moments. Equilibration of a triblock copolymer system from
a disordered initial configuration to a hexagonal phase is examined
using the algorithm.
\end{abstract}

\section{\label{sec:Introduction}Introduction}

Structures obtained from surfactant solutions or block copolymers
and their properties are of interest in many fields \cite{Smart2008}.
The processes that occur during these transitions can be relevant
in polymer processing for the fabrication of nanostructured materials
\cite{Klok2001}, especially in cases where the reversibility between
structures is of interest. In biophysics, the lamellar to reverse
hexagonal phase transition is considered as the first step in understanding
membrane fusion \cite{Siegel97,May2000,Koltover98}. 

Order-disorder and order-order transitions have been studied experimentally,
theoretically and using computer simulations. Much of the theoretical
and simulation work has focused on the identification of different
ordered structures that can be obtained when changing the architecture
of the amphiphilic molecule and the system conditions \cite{Ortiz2006}.
Nevertheless, the dynamic processes that describe the order-disorder
or order-order transitions are also important. Processes such as micelle
formation and stabilization, sphere to rod transitions, bilayer breakdown
and structural changes during the formation of nanostructured materials
have attracted much attention \cite{Gradzielski2004,Sakurai1993}.

Groot and Madden used dissipative particle dynamics (DPD) simulations
to describe the formation of an hexagonal phase from a disordered
phase, where an unstable gyroid phase appears as an intermediate using
DPD simulations \cite{Groot1998}. More recently, Soto-Figueroa et
al. described the dynamics of different order-order transitions in
polystyrene-polyisoprene diblock copolymers \cite{Figueroa2008}.
The transition between hexagonally packed cylinders to an array of
body centred cube spheres is a result of undulations in the cylinders
that eventually break into ellipsoids to latter form spheres. They
also describe the transition from a bicontinuous structure to a lamellar
phase going through an intermediate phase containing infinite cylinders,
not observed by Groot and Madden, before a lamellar phase is obtained.
Dynamics of the formation of ordered phases has also been reported
for a variety of surfactant architectures \cite{Khokhlov2008}, but
most of the results are limited to a collection of snapshots at different
times during the simulation. In some cases, an order parameter is
defined and used to determine the evolution of the observed phases,
which requires the calculation of the structure factor \cite{Groot1999}.

For any simulation approach which seeks to model mesoscale aggregates
from the microscale, the aggregates will appear infinite, that is,
they will percolate, spanning the periodic boundary conditions (PBCs)
of the simulation. The dimensionality of the aggregate (whether one
dimensional for cylinders, two dimensional for lamellae, or three
dimensional for network structures) will set the dimensionality of
this percolation. Algorithms exist to identify aggregates within a
simulation configuration (principally, the Hoshen-Kopelman (H-K) algorithm
\cite{hoshenkopelman}), however, once identified, the aggregate must
be properly characterised. 

The standard approach to characterisation of the size and shape of
an aggregate is diagonalization of its gyration tensor $\mathbf{S}$.
If the positions of the $N$ particles in the aggregate relative to
its center of mass are given by $\left\{ \mathbf{r}_{i}\right\} $,
the gyration tensor is given by:

\begin{equation}
S_{\alpha\beta}=\frac{1}{N}\sum_{i=1}^{N}r_{i}^{\alpha}r_{i}^{\beta}\label{eq:gyrationtensor}\end{equation}

\noindent where $r_{i}^{\alpha}$ denotes the $\alpha$'th component
of the vector $\mathbf{r}$, and $S_{\alpha\beta}$ indicates the
$\left(\alpha,\beta\right)$'th element of the tensor $\mathbf{S}$.
The eigenvectors of this tensor (the {}``principal axes'') give
the orientation of the aggregate, and the associated eigenvalues (the
{}``principle moments'') give the length scales of the aggregate
along these vectors.

\begin{figure}
\begin{centering}
\includegraphics{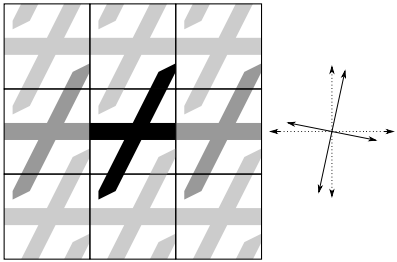}
\par\end{centering}

\caption{\label{fig:Pathological} An aggregate (filled area) repeated across
periodic boundary conditions. The micro-scale aggregate identified
from the structure is shown by the black filled area, and the gyration
tensor for this aggregate is schematically illustrated by the full
arrows. To calculate the gyration tensor for the true percolating
aggregate, only particles in the black and dark shaded areas should
be included. The gyration tensor for the percolating aggregate is
schematically illustrated by the dotted arrows. The two gyration tensors
are not aligned. Note that the percolating aggregate is infinite in
the horizontal direction, whereas the longest principle component
of the microscale aggregate lies closest to the vertical direction.}

\end{figure}

A naive implementation of Eqn. \ref{eq:gyrationtensor} for a percolating
aggregate, using only the coordinates from the \emph{microscale} aggregate
identified within the simulation boundary conditions, will produce
incorrect results. First, such an implementation will not identify
the correct principle moments; since the full, \emph{macroscale }aggregate
is percolating, at least one principle moment is infinite by definition.
Second, there is some subtlety as to which periodic images of a particle
to include in the calculation, such that only particles within a \emph{single}
aggregate should be included, and particles within \emph{images} of
the aggregate must be excluded. Finally, by not explicitly dealing
with the percolation of the aggregate, pathological configurations
may result in an incorrectly oriented gyration tensor. If the principle
axes are incorrectly oriented, then equivalently the principle moments
will be wrong. This is illustrated in \ref{fig:Pathological}.%

In this work, we present an algorithm (Calculation of Size of Percolating
Aggregates, or CaSPA) to deal with these issues. The algorithm is
demonstrated through application to aggregates formed in a triblock
copolymer system, simulated using dissipative particle dynamics (DPD)
\cite{DPD1,DPD2,FrenkelSmit,Groot}.

\section{\label{sec:Methods}Methods}

\subsection{\label{sub:CASPA}CaSPA algorithm}

Consider an aggregate composed of $N$ particles $\left\{ \mathbf{r}_{i}\right\} $,
with gyration tensor $\mathbf{S}$. The aggregate exists within PBCs,
and is a percolating cluster, that is, particles within the aggregate
are connected to particles in certain periodic images of the aggregate.
We denote the set of translation vectors between the contacting images
$\left\{ \mathbf{p}'_{i}\right\} $; these can be reduced to a combination
of $n_{p}$ linearly independent vectors $\left\{ \mathbf{p}_{i}\right\} $.
The aggregate is therefore percolating in $n_{p}$ dimensions. We
seek the gyration tensor $\mathbf{S}^{macro}$ for the full \emph{macroscale
}aggregate, made up of the combination of individual, contacting aggregates
across the periodic boundaries. If we include $n_{i}$ image aggregates
along each of the $\mathbf{p}$ vectors, this gyration tensor will
be given by:

\begin{eqnarray}
S_{\alpha\beta}^{macro} & = & \frac{1}{Nn_{i}^{n_{p}}}\left(\sum_{k_{1}=-(n_{i}-1)/2}^{(n_{i}-1)/2}\dots\sum_{k_{n_{p}}=-(n_{i}-1)/2}^{(n_{i}-1)/2}\right)\times\nonumber \\
 &  & \sum_{i=1}^{N}\left(r_{i}^{\alpha}+\sum_{j=1}^{n_{p}}k_{j}p_{j}^{\alpha}\right)\left(r_{i}^{\beta}+\sum_{j'=1}^{n_{p}}k_{j'}p_{j'}^{\beta}\right)\label{eq:fullmacroS}\end{eqnarray}

Multiplying out, through the symmetry of the sum limits, and considering
that since the origin is at the center of mass $\sum_{i}p_{i}^{\alpha}=0$,
we find:

\begin{equation}
S_{\alpha\beta}^{macro}=S_{\alpha\beta}+\frac{1}{12}\left(n_{i}-1\right)\left(n_{i}+1\right)Q_{\alpha\beta}\label{eq:macroS}\end{equation}

\noindent where the tensor $\mathbf{Q}$ is given by:

\noindent \begin{equation}
Q_{\alpha\beta}=\sum_{i=1}^{n_{p}}p_{i}^{\alpha}p_{i}^{\beta}\label{eq:Qtensor}\end{equation}
Note the similarity between the form of $\mathbf{Q}$ and the definition
of the gyration tensor (Eq. \ref{eq:gyrationtensor}). The tensor
$\mathbf{Q}$ can be considered as the normalized gyration tensor
for a coarse-grained representation of the macroscale aggregate, with
one point mass per image. We seek to diagonalise $\mathbf{S}^{macro}$,
giving three eigenvectors (the principle axes) $\mathbf{A}_{i}$,
and three corresponding eigenvalues (the principle components) $\lambda_{i}^{2}$,
$\lambda_{1}^{2}\leq\lambda_{2}^{2}\leq\lambda_{3}^{2}$.

As $n_{i}$ tends to infinity (the bulk limit), the gyration tensor
$\mathbf{S}^{macro}$ becomes dominated by the matrix $\mathbf{Q}$.
This determines the orientation of the macroscale aggregate in space.
However, provided the aggregate does not percolate in all dimensions,
one or more of the eigen\emph{values} of $\mathbf{Q}$ will be zero,
hence $\mathbf{Q}$ is singular. By performing \emph{singular value
decomposition} \cite{NRC} on $\mathbf{Q}$, we can identify the \emph{range}
$\mathbf{U}$ of $\mathbf{Q}$, the set of $n_{p}$ eigenvectors with
non-zero eigenvalues, and the \emph{nullspace} $\mathbf{V}$ of $\mathbf{Q}$,
a set of orthogonal eigenvectors for which the eigenvalues are zero.
The macroscale aggregate will have infinite principle components in
the range (where $\mathbf{Q}$ will dominate), and finite principle
components in the nullspace (where only $\mathbf{S}$ contributes).

For tensors $\mathbf{Q}$ with a single vector $\mathbf{V}_{1}$ in
the nullspace, the orientation of the finite principle component of
the macroscale aggregate is given by $\mathbf{A}_{1}=\mathbf{V}_{1}$
and the value of the finite principal component is given by the projection
of $\mathbf{S}$ along this vector, $\lambda_{1}^{2}=\mathbf{V}_{1}^{T}\mathbf{S}\mathbf{V}_{1}$.
The remaining two principle components are infinite, $\lambda_{2}^{2}=\lambda_{3}^{2}=\infty$,
with the corresponding principle axes lying along the range, $\mathbf{A}_{2}=\mathbf{U}_{2}$
and $\mathbf{A}_{3}=\mathbf{U}_{3}$.

For a tensor $\mathbf{Q}$ with two vectors ($\mathbf{V}_{1}$ and
$\mathbf{V}_{2}$) in the nullspace, the orientation of the infinite
principle component $\lambda_{3}^{2}=\infty$ of the macroscale aggregate
is given by $\mathbf{A}_{3}=\mathbf{U}_{3}$. The vectors describing
the finite principal components will be linear combinations of $\mathbf{V}_{1}$
and $\mathbf{V}_{2}$, which will be a pair of vectors which are orthogonal
to the range $\mathbf{U}_{3}$, but are otherwise arbitrary. To find
the finite principle axes, we must project the matrix $\mathbf{S}$
into the plane given by $\mathbf{V}$. This results in a two dimensional
gyration tensor given by $\left[\begin{array}{c}
\mathbf{V}_{1}^{T}\\
\mathbf{V}_{2}^{T}\end{array}\right]\mathbf{S}\left[\begin{array}{cc}
\mathbf{V}_{1} & \mathbf{V}_{2}\end{array}\right]$, with two-dimensional eigenvectors $\mathbf{A}'_{1}$ and $\mathbf{A}'_{2}$,
and corresponding eigenvalues giving the principle components, $\lambda_{1}^{2}$
and $\lambda_{2}^{2}$. The orientations of the principle components
are given by $\mathbf{A}_{i}=\mathbf{A}'_{i}\left[\mathbf{V}_{1}\,\mathbf{V}_{2}\right]$,
that is, the eigenvectors $\mathbf{A}'_{i}$ give the appropriate
linear combinations of the nullspace vectors to give the principle
axes.

The two remaining cases are tensors $\mathbf{Q}$ with no nullspace,
representing a percolating macroscale aggregate which is infinite
in all directions ($\lambda_{1}^{2}=\lambda_{2}^{2}=\lambda_{3}^{2}=\infty$),
and tensors $\mathbf{Q}$ with a three-dimensional nullspace, representing
aggregates with no self contacts, which can be treated entirely from
the aggregate gyration tensor $\mathbf{S}$.

From this, the algorithm to calculate the principle components and
axes of the macroscale aggregate must perform the following steps:

\begin{enumerate}
\item Identify the aggregate coordinates $\left\{ \mathbf{r}_{i}\right\} $,
and calculate the aggregate gyration tensor $\mathbf{S}$.
\item Identify the set of self-contact vectors, $\left\{ \mathbf{p}'_{i}\right\} $,
and reduce it to a linearly dependent set, $\left\{ \mathbf{p}_{i}\right\} $.
\item Calculate the tensor $\mathbf{Q}\left(\left\{ \mathbf{p}_{i}\right\} \right)$,
and find its range $\mathbf{U}$ and nullspace $\mathbf{V}$.
\item Calculate the projection of the gyration tensor $\mathbf{S}$ onto
the nullspace $\mathbf{V}$.
\item Return the macroscale aggregate principle axes (the range $\mathbf{U}$
and unit vector projections of the nullspace $\mathbf{V}$ onto the
gyration tensor $\mathbf{S}$) and principle components (infinity
for axes corresponding to the range, and the length of the vector
projections of the nullspace $\mathbf{V}$ onto the gyration tensor
$\mathbf{S}$ for the remaining components).
\end{enumerate}
We now describe how such an algorithm may be implemented.

\subsubsection{\label{sub:stitch}Identification of the aggregate coordinates}

\begin{figure}

\begin{centering}
\includegraphics{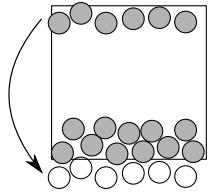}
\par\end{centering}

\caption{\label{fig:stitchpic}Illustration of the {}``stitching'' process.
Shaded circles represent particles within the PBC's which form two
disjunct segments of a single aggregate. Once self-contact vectors
have been identified, one of these segments is translated outside
of the PBC's (open circles) such that the aggregate forms a single,
connected cluster.}

\end{figure}

The H-K algorithm is suitable for initial identification of an aggregate
from a particle configuration. However, to deal with an aggregate
which crosses PBC's, an extra step is necessary. In simulation coordinates,
aggregates will often consist of a number of disjunct segments, connected
across the PBC's (see \ref{fig:stitchpic}). The first task of the
algorithm, once the initial $N$-particle aggregate has been identified,
is to {}``stitch'' these disjunct segments into a single, fully
connected object. The minimum image convention cannot be used for
this, since the positions of particles in \emph{images }of the aggregate
may become mixed with positions of particles in the aggregate during
calculation of the gyration tensor, giving incorrect results.

\begin{figure*}
\begin{centering}
\includegraphics{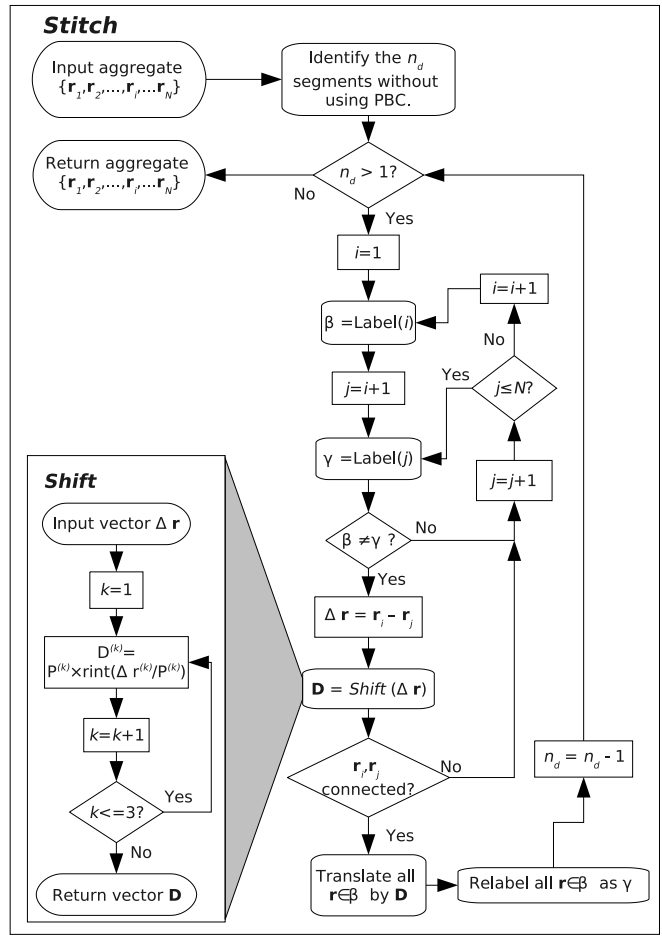}
\par\end{centering}

\caption{\label{fig:stitch}The {}``stitching'' algorithm. Inset: the {}``shift''
algorithm, which finds the vector between periodic images. The function
rint rounds its argument to the nearest integer. The vector $\mathbf{P}$
contains the lengths of the periodic boundary conditions.}

\end{figure*}

To {}``stitch'' an aggregate together, the coordinates $\left\{ \mathbf{r}_{i}\right\} $
of the particles in the aggregate are passed to a version of the cluster
identification algorithm which does \emph{not }recognise the PBCs.
This identifies the $n_{d}$ disjoint segments, and returns a list
$\left\{ l_{i}\right\} $, labelling each of the $N$ particles in
$\left\{ \mathbf{r}_{i}\right\} $ according to which subcluster that
particle is a member of. The algorithm then loops over all pairs of
particles, until it finds a pair $\left(i,j\right)$ such that $l_{i}\neq l_{j}$,
and which contact \emph{across} PBCs. Having found such a pair, all
particles with label $l_{i}$ are relabelled with $l_{j}$ and translated
such that particles $i$ and $j$ are bonded \emph{without }PBCs (merging
disjunct clusters $l_{i}$ and $l_{j}$ into a single, connected cluster),
and $n_{d}$ is decremented. This is then repeated until $n_{d}=1$,
that is, there exists only a single connected cluster. Note that the
resulting set of coordinates will now extend outside of the original
PBCs%
 (see \ref{fig:stitchpic}). A flowchart for the {}``stitching''
algorithm is shown in Fig. \ref{fig:stitch}%
. Given this single connected cluster (the microscale aggregate),
the gyration tensor $\mathbf{S}$ can be calculated as Eqn. \ref{eq:gyrationtensor}.

\subsubsection{\label{sub:self-contact}Identification of the Self-Contact Vectors}

Once the aggregate coordinates have been identified, the self-contact
vectors can be found. The set $\left\{ \mathbf{p}'_{i}\right\} $
is easy to identify - loop across every pair of particles in the aggregate,
check if they are contacting across the PBCs, and if they are, store
the vector connecting the periodic images (given, in the terminology
of \ref{fig:stitch}, by \emph{Shift$\left(\Delta r\right)$}). This
set may, however, be very large, and will in general contain a large
quantity of redundant data. We seek instead the linearly independent
set $\left\{ \mathbf{p}_{i}\right\} $, which in three dimensions
may consist of a maximum of three vectors. 

To find these vectors, the algorithm must loop across every pair of
particles in the aggregate, checking for contact across PBCs, until
a vector connecting periodic images is found. This is stored in $\mathbf{p}_{1}$.
The loop then continues, until a second candidate vector $\mathbf{p}'$
is found. The algorithm must check that this new vector is linearly
independent of $\mathbf{p}_{1}$; the condition for this is that $\left|\mathbf{p}_{1}\times\mathbf{p}'\right|\neq0$.
If $\mathbf{p}'$ is linearly independent of $\mathbf{p}_{1}$, then
the algorithm sets $\mathbf{p}_{2}=\mathbf{p}'$. Otherwise, the algorithm
continues to check candidate vectors $\mathbf{p}'$ for linear dependence
against $\mathbf{p}_{1}$.

If two linearly dependent vectors have been found, then the loop once
again continues, but now candidate vectors must be checked for linear
independence against $\mathbf{p}_{1}$ and $\mathbf{p}_{2}$. The
condition for linear independence is now that $\det\left(\left[\mathbf{p}_{1}\,\mathbf{p}_{2}\,\mathbf{p}'\right]\right)\neq0$.
If $\mathbf{p}'$ is linearly independent of $\mathbf{p}_{1}$, the
algorithm sets $\mathbf{p}_{3}=\mathbf{p}'$, and terminates; the
aggregate is percolating in three dimensions, hence all principle
components are infinite. 

\begin{figure*}
\begin{centering}
\includegraphics{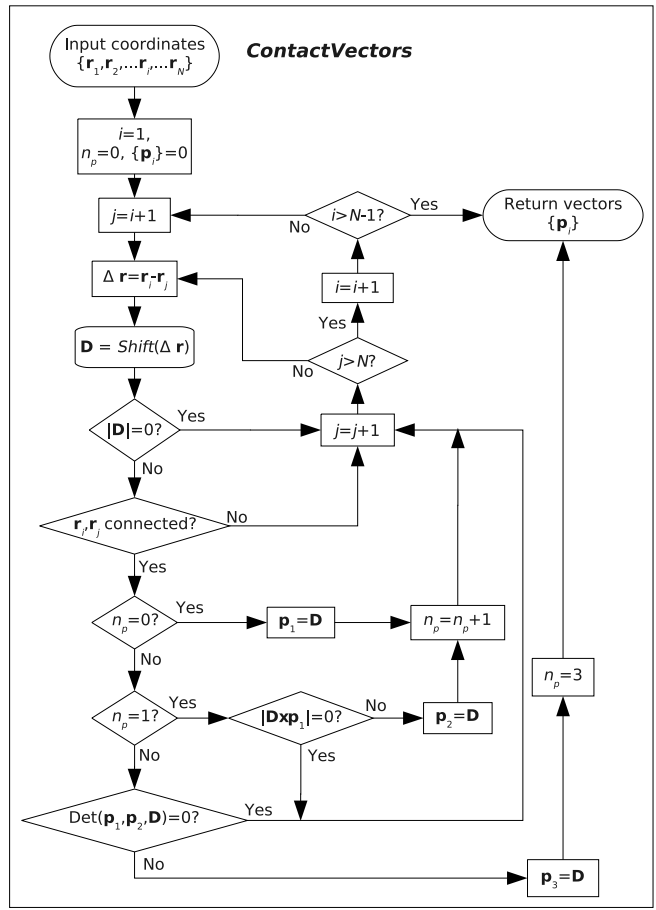}
\par\end{centering}

\caption{\label{fig:contactvectors}The ContactVectors algorithm.}

\end{figure*}

Once all pairs have been checked, the algorithm is left with $n_{p}$
linearly independent self-contact vectors $\left\{ \mathbf{p}_{i}\right\} $.
A flowchart for the self-contact identification algorithm is shown
in \ref{fig:contactvectors}.%
From the vectors $\left\{ \mathbf{p}_{i}\right\} $, the tensor $\mathbf{Q}$
may be calculated as Eqn. \ref{eq:Qtensor}. This tensor is then diagonalised.
There should be $n_{p}$ non-zero eigenvalues, corresponding to the
eigenvectors $\left\{ \mathbf{U}_{i}\right\} $ in the range of $\mathbf{Q}$,
and $\left(3-n_{p}\right)$ zero eigenvalues, corresponding to the
eigenvectors $\left\{ \mathbf{V}_{i}\right\} $ in the nullspace of
$\mathbf{Q}$.

\subsubsection{\label{sub:Gyration-Tensor}Calculation of Gyration Tensor of the
Infinitely Repeated Aggregate}

\begin{figure*}
\begin{centering}
\includegraphics{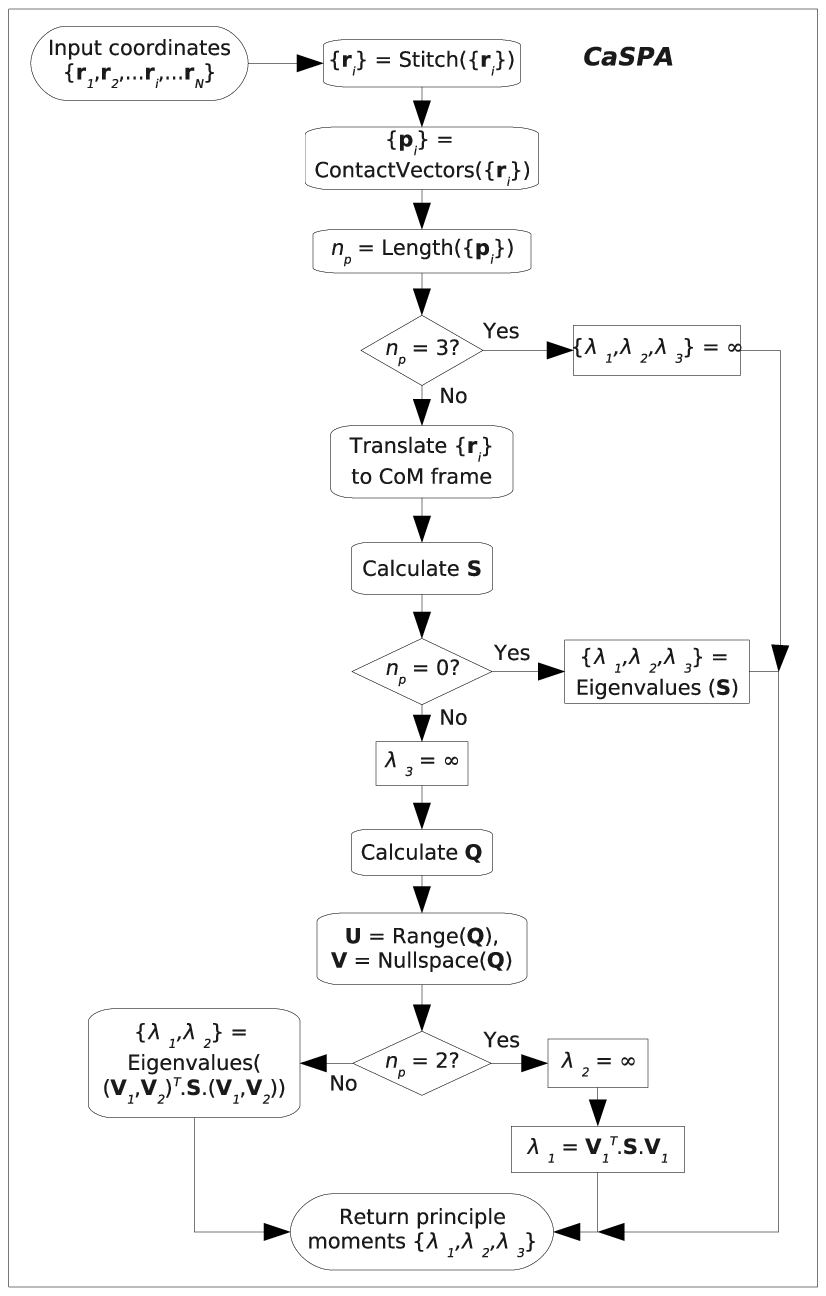}
\par\end{centering}

\caption{\label{fig:CaSPA}The CaSPA algorithm; {}``CoM frame'' denotes the
frame of reference where the aggregate center of mass is the origin. }

\end{figure*}

Finally, the algorithm has the aggregate gyration tensor $\mathbf{S}$,
the dimensionality of percolation, $n_{p}$, and the range and nullspace
of percolation, $\left\{ \mathbf{U}_{i}\right\} $ and $\left\{ \mathbf{V}_{i}\right\} $.
The algorithm will already have terminated if the aggregate is percolating
in three dimensions, returning three infinite principle components.
Otherwise, the principle components and axes may be calculated by
projection of $\mathbf{S}$ onto the nullspace $\mathbf{V}$, as detailed
above. A full flowchart of CaSPA is shown in \ref{fig:CaSPA}.%

\subsubsection{\label{sub:Optimization}Optimization}

The algorithm as described here has been separated into parts for
conceptual clarity, and as such is not optimized for speed. It is
anticipated, however, that the generation of uncorrelated configurations
by simulation for analysis is likely to be significantly slower than
the analysis of these configurations.

The process of checking all pairs in the system means that, as presented,
the time needed scales as $N^{2}$. If the definition of contact allows,
the algorithm can be reduced to scaling with $N$ via construction
and use of a cell list. A second obvious improvement would be to perform
the identification of self-contact vectors $\left\{ \mathbf{p}_{i}\right\} $
at the same time as the {}``stitching'' algorithm, by checking for
boundary-crossing contacts between particles in the same aggregate
segment. While this will not change the scaling form, it should significantly
reduce the constant of proportionality.

\subsection{\label{sub:triblock}Test System}

To demonstrate the algorithm, we have simulated equilibration of a
triblock copolymer solution from an initial, random configuration.
The simulation has been performed using dissipative particle dynamics
(DPD). This is a mesoscale simulation methodology, where {}``particles''
represent loosely-defined {}``fluid elements'', and move via a combination
of dissipative, random and conservative interparticle forces. 

Particles in the system interact via the {}``standard'' DPD force
field \cite{FrenkelSmit}, given by:

\begin{equation}
U_{ij}=\begin{cases}
\frac{a_{ij}}{2r_{c}}\left(r_{c}-r_{ij}\right)^{2} & r_{ij}\leq r_{c}\\
0 & r_{ij}>r_{c}\end{cases}\label{eq:DPDlabel}\end{equation}

\noindent where $r_{ij}$ is the separation between particles indices
$i$ and $j$, $r_{c}$ is a notional particle diameter (setting the
length scale of the system), and $a_{ij}$ a parameter which depends
upon the types of particles indices $i$ and $j$, setting the strength
of interaction between these types. Particles which are bonded along
a polymer chain also interact via a spring force, spring constant
$C=4k_{B}T$ \cite{Groot}.

The simulated system exists within a cubic box of size $15r_{c}$,
with total particle number density $\rho=3$, and is made up of $A$
and $B$ type monomers and solvent particles (type $S$). Interaction
parameters for like particle types are $a_{XX}=25$, and for unlike
types are $a_{SA}=a_{AB}=50$ and $a_{SB}=25$; as such, $A$-type
monomers are {}``hydrophobic''. All $A$ and $B$ type monomers
exist as part of $A_{3}B_{10}A_{3}$ triblock copolymers. The system
has $4061$ solvent particles, and $379$ copolymers, giving a copolymer
volume fraction of $0.6$. The simulation reduced timestep is $\Delta t=0.01$,
and the simulation is run for $4\times10^{6}$ timesteps. The initial
configuration is generated by placing solvent particles and one end
monomer of each copolymer randomly in the simulation box. Copolymers
are grown by placing the center of the next monomer in each polymer
at a random position on a sphere of radius $r_{c}$ centered on the
previous monomer. The energy of this random configuration is then
minimised by steepest descent to give the starting configuration for
the simulation.

\section{\label{sec:Results}Results}

\begin{figure}
\begin{centering}
\includegraphics{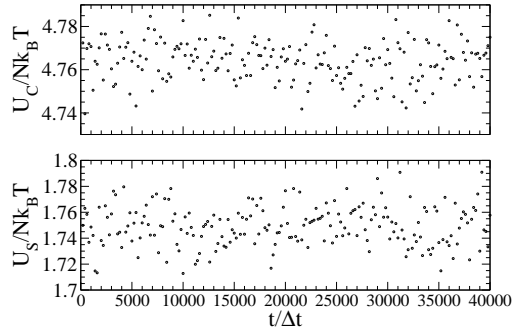}
\par\end{centering}

\caption{\label{fig:triblockU}Configurational energy for the triblock copolymer
simulation, decomposed into pair interactions ($U_{C}$) and spring
force interactions ($U_{S}$), as a function of simulation time $t$.
No significant energetic relaxation is observed.}

\end{figure}

Time series data of the configurational energy (decomposed into pair
and spring interactions) from the simulation are shown in \ref{fig:triblockU}.
From the point of view of energy, the random initial system configuration
appears to reflect the equilibrium state well, and the system does
not appear to pass through any significant energetic relaxation.%

\begin{figure}
\begin{centering}
\includegraphics{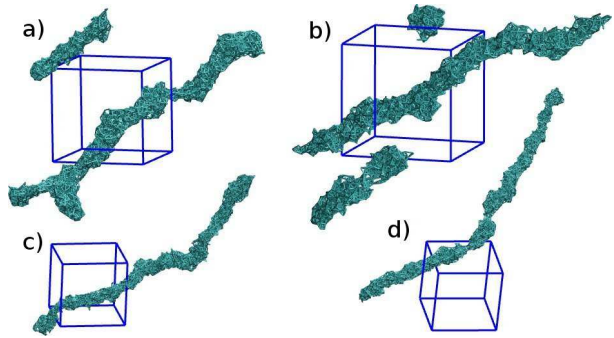}
\par\end{centering}

\caption{\label{fig:snapshots}''Stitched'' snapshot conformations from the
triblock copolymer simulation. Cylinders join $A$-type monomers separated
by less than $r_{c}$. Lines indicate the simulation box. a) Coexistence
between large, self-contacting aggregate (lower right) and elongated
micelle at $t=2\Delta t$. b) Coexistence between large, self-contacting
aggregate (center), spherical micelle (top) and elongated micelle
(bottom) at $t=100\Delta t$. c) Single, elongated micelle at $t=200\Delta t$.
d) Large, self-contacting aggregate at $t=380\Delta t$.}

\end{figure}

{}``Stitched'' snapshot configurations from the simulation (see
\ref{fig:snapshots})%
 indicate that the system undergoes mesophase separation, with the
{}``hydrophobic'' $A$-type monomers forming elongated aggregates,
demixed from the solvent and $B$-type monomers. As such, the CaSPA
algorithm has been used to study these aggregates. Pairs of $A$-type
monomers are considered to be {}``connected'' if they overlap (that
is, if $r_{ij}<r_{c}$). Data for clusters smaller than 3 monomers
are discarded. It should be noted that the results we present are
insensitive to the definition of connectivity, being qualitatively
similar across a range of definitions of connectivity from $r<0.8r_{c}$
to $r<1.2r_{c}$. This insensitivity can be explained by considering
the radial distribution function between $A$-type monomers $g_{AA}(r)$%
\begin{figure}
\begin{centering}
\includegraphics{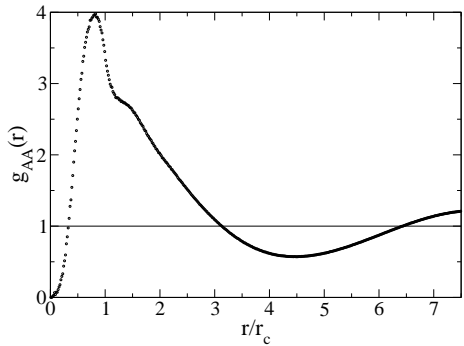}
\par\end{centering}

\caption{\label{fig:triblockgr}Radial distribution function $g_{AA}(r)$ between
hydrophobic monomers.}

\end{figure}
 (see \ref{fig:triblockgr}). The nearest-neighbour peak can be seen
to be at a maximum at at $r=0.825r_{c}$, and can further be seen
to be distinct from the broader peak indicating aggregate structure
up to $r\approx1.2r_{c}$. As long as the cutoff for connectivity
lies between these two values, the algorithm will probe nearest neighbour
structure in an effective fashion.

\begin{figure}
\begin{centering}
\includegraphics{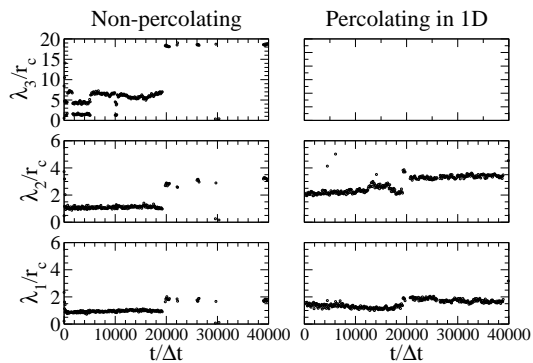}
\par\end{centering}

\caption{\label{fig:triblockrgs}Principal components of the gyration tensor
$\lambda_{1}\leq\lambda_{2}\leq\lambda_{3}$ for clusters of $A$-type
monomers as a function of simulation time $t$. The left hand column
shows principal components for non-percolating clusters. The right
hand column shows principal components for clusters which are percolating
in one dimension; note that, for these clusters, $\lambda_{3}=\infty$
by definition. Multiple points at a given time step denote multiple
clusters of that type at a given time step.}

\end{figure}

\begin{figure}
\begin{centering}
\includegraphics{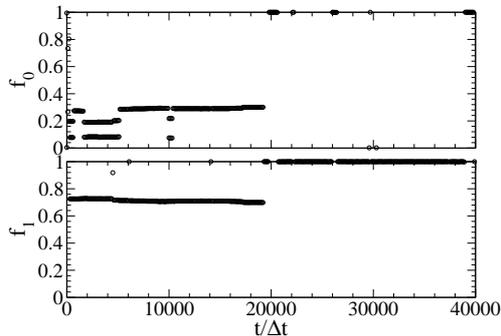}
\par\end{centering}

\caption{\label{fig:frac}Fraction of $A$-type monomers in non-percolating
clusters ($f_{0}$) and in clusters which are percolating in one dimension
($f_{1}$) as a function of simulation time, $t$. Multiple points
at a given time step indicate multiple clusters of that type at that
time step.}

\end{figure}

Time series data from CaSPA for the principal gyration components
of these clusters are shown in \ref{fig:triblockrgs}. Within these
results, clusters of $A$-type monomers are found to be either non-percolating,
or to be percolating in only one dimension. Time series data for the
fraction of $A$-type monomers in clusters of each type are shown
in \ref{fig:frac}. While no obvious energetic relaxation is observed,
the CaSPA results suggest significant structural relaxation. %

The CaSPA data clearly shows that, across most of the simulation,
the system is dominated by the presence of a single, large (containing
more than $70\%$ of all $A$-type monomers) cluster which is percolating
in one dimension. As such, we can suggest that the system is equilibrating
towards a hexagonal cylindrical liquid crystal. 

Up until $t\approx19300\Delta t$, the percolating cluster coexists
with one or more smaller, though still significant, non-percolating
clusters, which appear to be micellar in character. Where only a single
micelle exists, it can be seen to be extended, and approximately cylindrical
($\lambda_{1}\approx\lambda_{2}<\lambda_{3}$) (see snapshot configuration
in \ref{fig:snapshots}(a)). The largest principle component of this
micelle can be seen to be of the order of half the length of the simulation
box ($\lambda_{3}\approx7r_{c}$). Where two micelles exist, the smaller
micelle is approximately spherical ($\lambda_{1}\approx\lambda_{2}\approx\lambda_{3}\approx r_{c}$),
while the larger micelle is extended and approximately cylindrical
(see snapshot configuration in \ref{fig:snapshots}(b)). The excess
material contained in these micelles is absorbed into the larger cluster
at $t\approx19300\Delta t$. The smallest principal moments of gyration
$\lambda_{1}$ and $\lambda_{2}$ increase at around $t\approx19300\Delta t$,
corresponding to absorption of this material into the large micelle.

After $t\approx19300\Delta t$, the vast majority (greater than $99.5\%$)
of $A$-type monomers exist in a single cluster. Across most of this
time (approximately $85\%$), that cluster is percolating in one dimension
(see snapshot configuration in \ref{fig:snapshots}(d)), however,
there are five events where that cluster becomes non-percolating (see,
for e.g., snapshot configuration in \ref{fig:snapshots}(c)). During
these events, the cluster keeps approximately the same values for
the smaller principal moments of gyration, while taking on a large
value ($\lambda_{3}\approx18.5r_{c}$) for the largest principal moment. 

The character of the large, percolating cluster does not appear quite
cylindrical from the data; the aspect ratio of the cluster is not
unity, but instead $\lambda_{2}/\lambda_{1}\approx1.9$. This value
does not change significantly with time. The reason for this apparent
non-cylindrical character is made clear in the {}``stitched'' snapshot
configurations (\ref{fig:snapshots}); the cluster does not lie along
a straight path, but is instead significantly curved. This suggests
that the configuration is not fully equilibrated, but is suffering
from a finite size effect, where the cylindrical aggregate is not
aligned properly within the simulation box, and must bend in order
to maintain self-contact across periodic boundaries. We suggest that
the observed events where the cluster becomes non-percolating correspond
to equilibration towards the correct alignment.

\section{\label{sub:Discussion}Discussion}

In the previous sections, we have presented CaSPA, a novel algorithm
to characterise the size and orientation of percolating aggregates,
and have tested this algorithm on a triblock copolymer system. Results
using CaSPA show that this system forms percolating one-dimensional
aggregates. These results demonstrate the strengths of the algorithm.
Configurational energy does not indicate any relaxation phenomena
during the simulation (\ref{fig:triblockU}), whereas measurement
of the size of aggregated clusters clearly identifies structural relaxation
(\ref{fig:triblockrgs}). While the standard Hoshen-Kopelman algorithm
would identify the presence of this relaxation, CaSPA allows a fuller
description by identifying percolating aggregates. In the case presented
here, this allows very long elongated micelles to be distinguished
from true percolating cylindrical aggregates. Further, CaSPA returns
the correct principle moments of gyration for branched percolating
structures (such as the percolating aggregates in \ref{fig:snapshots}
(a) and (b)). Finally, the {}``stitching'' step in CaSPA can be
used to generate snapshot configurations of aggregates (\ref{fig:snapshots})
which are easier to visualise than the equivalent {}``unstitched''
configurations. Coupled with data on the size of aggregates, this
clearer visualization can aid both understanding and communication
of observed behaviors in simulation of mesostructured materials.

Use of CaSPA is complementary to other order parameter measurements
used to describe mesophase structures. While the algorithm provides
a broad range of information on the size and shape of aggregates,
it does not provide information on the degree of segregation of aggregating
particles from the rest of the system. As a comparative example, the
$P$ order parameter proposed by Groot et al \cite{Groot1999} provides
information on the degree of segregation of particles, and can differentiate
between structures, but is not intended to provide clear geometric
information on those structures. In summary, CaSPA provides a useful
addition to methods for identification and characterisation of mesostructured
materials.

\bibliographystyle{prsty}
\bibliography{clustersizeanalysis}

\end{document}